\documentclass[aps,pra,reprint,superscriptaddress,nofootinbib,showkeys]{revtex4-2} %linenumbers preprint reprint
\usepackage[english]{babel}
\usepackage{amsmath,amssymb,xfrac,mathtools}
\usepackage{hyperref}
\usepackage{graphicx,xcolor}
\usepackage{bm}

\let\oldhat\hat
\renewcommand{\hat}[1]{\oldhat{#1}}
\newcommand{\ex}{\mathrm{e}}
\newcommand{\ie}{\textit{i}.\textit{e}.\ }
\newcommand{\eg}{\textit{e}.\textit{g}.\ }

\begin{document}
\title[Distinguishing quantum features in classical propagation]{Distinguishing quantum features in classical propagation}
\author{Kelvin Titimbo}
\email{titimbo@itp.ac.cn}
\affiliation{CAS Key Laboratory of Theoretical Physics, Institute of Theoretical Physics, Chinese Academy of Sciences, Beijing 100190, China}
\affiliation{Centro Brasileiro de Pesquisas Fisicas, Rua Xavier Sigaud 150, 22290-180, Rio de Janeiro, R.~J., Brazil.}
\author{Gabriel M. Lando}
\altaffiliation[Present address: ]{Max Planck Institute for the Physics of Complex Systems, N\"othnitzer Strasse 38, D-01187, Dresden, Germany.}
\affiliation{Centro Brasileiro de Pesquisas Fisicas, Rua Xavier Sigaud 150, 22290-180, Rio de Janeiro, R.~J., Brazil.}
\author{Alfredo M. Ozorio de Almeida}
%\email{ozorio@cbpf.br}
\affiliation{Centro Brasileiro de Pesquisas Fisicas, Rua Xavier Sigaud 150, 22290-180, Rio de Janeiro, R.~J., Brazil.}

\date{\today}

\begin{abstract}
The strictly classical propagation of an initial Wigner function, referred to as TWA or LSC-IVR, is considered to provide approximate averages, despite not being a true Wigner function: it does not represent a positive operator. 
We here show that its symplectic Fourier transform, the truncated chord approximation (TCA), coincides with the full semiclassical approximation to the evolved quantum characteristic function (or chord function) in a narrow neighbourhood of the origin of the dual chord phase space. 
Surprisingly, this small region accounts for purely quantum features, such as blind spots and local wave function correlations, as well as the expectation of observables with a close classical correspondence.
Direct numerical comparison of the TCA with exact quantum results verifies the semiclassical predictions for an initial coherent state evolving under the Kerr Hamiltonian.
The resulting clear criterion for any further features, which may be estimated by classical propagation, is that, within the chord representation, they are concentrated near the origin.
\end{abstract}


\maketitle

\section{Introduction}\label{sec:I}

Classical molecular dynamics remains a widely used method 
to describe complex molecular systems, in spite of their fundamental quantum identity \cite{milw2001,milw2012}. 
The justification of such a radical approximation is that the Weyl representation treats observables as ordinary real functions in phase space and their expectation is also evaluated classically: it is reduced to an integral with the Wigner function \cite{wig1932}. 
The latter is a full quantum representation of the density operator, but \eg for a coherent state, it is merely a Gaussian in phase space. 
By identifying this with an initial Liouville probability distribution in phase space and then propagating it classically, one generates some kind of approximation to the quantum evolved Wigner function.

The ease with which one can employ this \textit{truncated Wigner approximation} (TWA) \cite{hele1976}, which is also denoted LSC-IVR (linearized approximation to the semiclassical initial value representation) in some contexts \cite{milw2001,milw2012}, affords no clue to its true status.
Even though the full Moyal equation of motion for the Wigner function has the classical Liouville equation as its first term in a formal expansion in powers of Planck's constant, $\hbar$, it is the remainder that produces all the interfering oscillations that proliferate and eventually mask the classically evolved Gaussian.
In practice, there are applications where the TWA is experimentally verified to be effective \cite{milw2001}.
On the other hand, it has been proved that the operator, with the Weyl representation that coincides with the classically evolved Liouville distribution, is not positive; in other words, it is not legitimate to interpret it as a Wigner function, not even for a mixed state \cite{habjm2002, klisr2020}.
In some ways, the TWA is more akin to one of those hybrid creatures in ancient Greek mythology than an objective scientific concept.

A close analogy can be made to the \textit{ergodic Wigner function} of the \textit{Berry-Voros conjecture} \cite{vor1976,ber1977a}.
This classical microcanonical density, a Dirac $\delta$-function on the energy shell, was proposed as some sort of approximation to the full Wigner function for the eigenstate of a classically chaotic system.
Notwithstanding the proof that again the operator represented by this approximation is not positive \cite{bal1980,ozo2014}, it not only supplies satisfactory expectations of simple observables, but it even provides purely quantum information about the eigenstate.
Indeed, \textit{local wave-function correlations} can be evaluated \cite{ber1977a}, as well as the closest \textit{blind spots}, that is, the minimal dislocations of the state which render it exactly orthogonal to itself \cite{zamo2009,zamo2010}.

In the case of the ergodic Wigner function, clarification follows from the reinterpretation of its symplectic Fourier transform (SFT).
Given a classical probability distribution, its SFT the \textit{characteristic function} encapsulates very useful information, such as all the moments of the distribution through mere differentiation, but it plays a subordinate role.
In contrast, within quantum mechanics the SFT of a Wigner function, the \textit{quantum characteristic function} or simply the chord function, rivals the Wigner function itself as a representation of the density operator, along with the chord representation of arbitrary operators; expectations and correlations are evaluated as integrals in the Fourier variables: the dual \textit{chord phase space}.

It is through the SFT of the ergodic Wigner function that one unravels its paradoxical features: it coincides with the true chord function only near the origin, but this small region of chord space does contain quantum information which is masked within the Wigner function itself.
Since no claim needs to be made about the accuracy of the SFT in the outlying regions of chord space, it makes no sense to censure this \textit{local ergodic approximation} for its inability to guarantee positivity of some operator, which would be portrayed by the chord function as a whole.

The objective of the present paper is to show that an analogous result holds for the classical evolution of a coherent state, the TWA: its SFT, the \textit{truncated chord approximation} (TCA), coincides with the full quantum evolution of the initial chord function near the origin of the chord phase space, even though the operator that is so represented cannot be positive.

The basic definitions of the Wigner-Weyl representation are reviewed in the following section, together with the semiclassical (SC) approximation to the evolution of the Wigner function, which replaces the Moyal expansion by a propagator.
The crucial point is that such propagator depends generally on pairs of trajectories and it is only where the appropriate pair remains close together throughout the time interval that they can be approximated by a single trajectory.

The rash assumption that this closeness holds for all pairs of trajectories then furnishes the TWA in section\ref{sec:III}.
In contrast, this same condition can be satisfied selectively within the chord representation: it is here deduced from the full SC approximation that the SFT of the TWA, that is, the TCA is valid close to the chord origin, no matter how widely off the mark it can be in outlying regions of this dual phase space.

The numerical computation of features, which depend solely on the central region of the chord function, were performed for the evolution of an initial coherent state generated by the quartic Kerr Hamiltonian presented in section \ref{sec:IV}.
This choice allows for complete control, in spite of being a nonlinear classical system, heavy laden with caustics \cite{lanvi2019}.
Focus is first placed on the evolution of \textit{blind spots}, that is, those displacements of a state which render it orthogonal to itself.
These are computed directly from the chord function in section \ref{sec:V}.
In section \ref{sec:VI} it is shown that the expectation of observables, corresponding to smooth functions in phase space, also depends only on the central region of the chord function.
We then compare the classical approximation of the moments of position, $\left< \hat{q}^{n} \right>$, and momentum, $\left<  \hat{p}^{n} \right>$, to their exact quantum evolution for the Kerr system.
Another family of properties, the \textit{local wave function correlations} \cite{ber1977a}, were shown in \cite{zamzo2015} to depend only on the central region of the chord function.
In section \ref{sec:VII} we verify this feature for the Kerr evolution of an initial coherent state, which classically is a spiral elongating and tightening in time. 
Having removed to the Appendix \ref{appen} our interpretation of how the distance between the spiral windings separates the valid classical region of the TCA from the purely quantum scattering of the outer blind spots, our conclusions are summarized in section \ref{sec:conclu}.

\section{Definitions and Semiclassical Review} \label{sec:II}

An evolving quantum state $\left| \psi(t) \right>$ is represented in phase space $\mathbf{x} = (p,q)$, the momentum and the position coordinates respectively, by the time dependent Wigner function \cite{wig1932}:
\begin{equation}\label{eq:wigner}
W(\mathbf{x},t) = \int \frac{\mathrm{d} \tilde{q}}{\pi\hbar} \left< q - \sfrac{\tilde{q}}{2} | \psi(t) \right> \left< \psi(t) |  q+ \sfrac{\tilde{q}}{2} \right> \ex^{i\tilde{q}p/\hbar}  .
\end{equation}
Here we consider the simple case of a single degree of freedom, but all the formulae can be generalized for $N$ degrees of freedom, such that $\mathbf{x} = \left( p_{1}, \dots, p_{N}, q_{1},\dots,q_{N} \right)$.

The SFT of the Wigner function is a function in the dual phase space of chords $\bm{\xi} = \left( \xi_{p}, \xi_{q} \right)$,
\begin{equation}\label{eq:chord_FT}
\chi(\bm{\xi},t) = \int \frac{\mathrm{d}\mathbf{x}}{2\pi\hbar} \exp\left[ -\frac{i}{\hbar} \mathbf{x}\cdot \mathrm{J}\bm{\xi} \right] W(\mathbf{x},t) \, ,
\end{equation}
where
\begin{equation}
    \mathrm{J} = \left(  \begin{array}{r@{\quad}cr}
        0 & -1 \\
    	1 & 0
	\end{array}\right)
\end{equation}
is the standard \textit{symplectic matrix}. 
This would be just the characteristic function if $W(\mathbf{x},t)$, which is real but can be negative, were a classical probability density.
Thus, $\chi(\bm{\xi},t)$ is referred to as the \textit{chord function}.
An alternative construction,
\begin{equation}\label{eq:chord}
\chi(\bm{\xi},t) = \int \frac{\mathrm{d}\tilde{q}}{2\pi\hbar} \left< \tilde{q} + \sfrac{\xi_{q}}{2} | \psi(t) \right> \left< \psi(t) |  \tilde{q} - \sfrac{\xi_{q}}{2} \right> \ex^{-i\tilde{q}\xi_{p}/\hbar} \, ,
\end{equation}
emphasises the full equivalence of this complex representation to the Wigner function.

The wave-function for a coherent state $\left| \alpha \right>$, with $\alpha = (\alpha_{q} + i\alpha_{p})/\sqrt{2\hbar}$, is defined as \cite{leonhardt}
\begin{equation}
    \left<q | \alpha \right> = \left( \frac{1}{\pi\hbar} \right)^{\frac{1}{4}} \exp\left[ -\frac{1}{2\hbar}(q-\alpha_{q})^{2} + i \frac{\alpha_{p}}{\hbar} \left( q - \frac{\alpha_{q}}{2} \right) \right] \, ,
\end{equation}
where one identifies the expectations $\left< \hat{q} \right> = \alpha_{q}$ and $\left< \hat{p} \right> = \alpha_{p}$.
Its Wigner function is the phase space Gaussian centred on $\bm{\alpha}$,
\begin{equation}\label{eq:coh_wig}
    W_{\alpha}(\mathbf{x}) = \frac{1}{\pi\hbar} \ex^{-\left( \mathbf{x} - \bm{\alpha} \right)^{2}/\hbar} \, ,
\end{equation}
where the real vector $\bm{\alpha} = (\alpha_{p},\alpha_{q})$.
In contrast, the corresponding chord function expresses the location of the coherent state through its phase,
\begin{equation}\label{eq:coh_chord}
    \chi_{\alpha}(\bm{\xi}) = \frac{1}{2 \pi\hbar} \ex^{-i \bm{\alpha}\cdot \mathrm{J}\bm{\xi}/\hbar} \ex^{-\bm{\xi}^{2}/4\hbar} \, ,
\end{equation}
so that it is always concentrated on the origin, whatever the displacement, $\bm{\alpha}$, of the coherent state.

\begin{figure}
    \centering
    \includegraphics[width=0.90\linewidth]{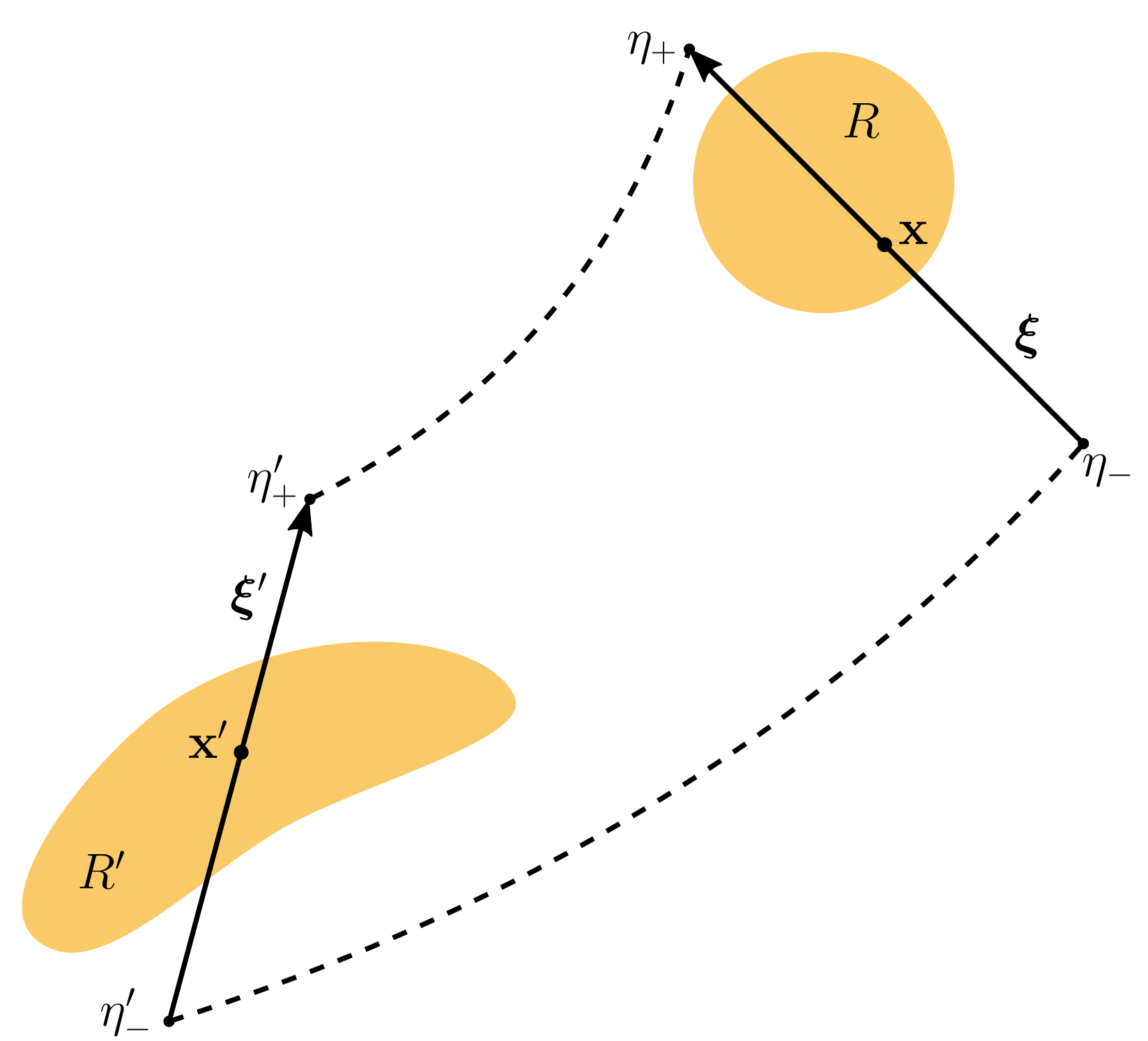}
    \caption{A final chord $\bm{\xi}'$, with centre $\mathbf{x}'$, is evolved backwards to form an initial chord $\bm{\xi}(\bm{\xi}',\mathbf{x}',t)$, centered at $\mathbf{x}(\bm{\xi}', \mathbf{x}', t)$. By considering the endpoints of $\bm{\xi}'$ and $\bm{\xi}$, such evolution is described by the circuit $\eta'_{-}\mapsto\eta_{-}\mapsto\eta_{+}\mapsto\eta'_{+}$. The first and last propagations are performed along classical trajectories (dashed lines), while the middle one is a reflection around $\mathbf{x}$, given by $\eta_{+} = 2\mathbf{x}- \eta_{-}$. The yellow areas $R$ and $R'$ represent the Wigner function at initial and final times, respectively.}
    \label{fig:traj}
\end{figure}

An arbitrary Hamiltonian, $\hat{H}$, evolves the coherent state according to
\begin{equation}\label{eq:evol}
    \left| \alpha \right>_{t} = \hat{U}_{t} \left| \alpha \right> = \ex^{-i t \hat{H}/\hbar} \left| \alpha \right> \, .
\end{equation}
The semiclassical approximation for its time evolved Wigner function, which was tested in \cite{lanvi2019}, depends on pairs of trajectories with initial values $\eta_{\pm}$ and final values $\eta'_{\pm}$ as depicted in Fig.~\ref{fig:traj}.
The initial centre and chord are given by $\mathbf{x} = (\eta_{+}+\eta_{-})/2$ and $\bm{\xi}=\eta_{+} - \eta_{-}$, respectively; the evolved centre and chord, $\mathbf{x}'$ and $\bm{\xi}'$, are defined accordingly from $\eta'_{\pm}$.
Then, the semiclassical evolution of the Wigner function employed in \cite{lanvi2019}, following \cite{ozovz2013}, is
\begin{multline}\label{eq:sc_wig}
W(\mathbf{x}', t) \approx \int \frac{\mathrm{d}\xi_{p}'\, \mathrm{d}\xi_{q}'}{2\pi\hbar} \left| \det \frac{\mathrm{d} \bm{\xi}}{\mathrm{d}\bm{\xi}'} \right|^{\frac{1}{2}} \\
\times \exp\left[ \frac{i}{\hbar} \left( S(\eta'_{\pm},\eta_{\pm}) + \mathbf{x}\cdot\mathrm{J}\bm{\xi} \right) \right] \chi(\bm{\xi}) \, ,
\end{multline}
where the initial chord $\bm{\xi}$, and the initial centre $\mathbf{x}$, are considered as dependent vectors on $\mathbf{x}'$, the argument of the evolved Wigner function, and the integration variables $\bm{\xi}'$.
Together they determine the endpoints, $\eta'_\pm$, which evolve backward to the initial points of the forward evolution, $\eta_\pm$.
The nonlinear action in the propagation kernel is
\begin{equation}\label{eq:action}
S(\eta'_{\pm},\eta_{\pm}) = t \bigl[ H(\eta_{+}) - H(\eta_{-}) \bigr] - \oint_{\mathcal{C}(\eta'_{\pm},\eta_{\pm})}^{}  p\, \mathrm{d}q \, .
\end{equation}
Here, $\mathcal{C}(\eta'_{\pm},\eta_{\pm})$ is the closed contour given by $\eta'_{-} \mapsto \eta_{-} \mapsto \eta_{+} \mapsto \eta'_{+}$, detailed in Fig.~\ref{fig:traj}.
We have here omitted any extra phase due to the Maslov index \cite{masf1981,lit1992,ozoi2014}, since its contribution can be neglected within a narrow region of the origin in the dual chord phase space and the short evolution times contemplated here.

\section{TWA and the Local Classical Approximation for the Chord Function}\label{sec:III}

No loss is incurred by the chord function for a coherent state \eqref{eq:coh_chord} by restricting its argument to small chords, but the final chords $\bm{\xi}'$ may achieve classical dimensions in a finite time.
Even so, sufficiently small initial chords do remain small for any given time and for these it is legitimate to approximate the nonlinear action \eqref{eq:action} as $S(\eta'_{\pm},\eta_{\pm}) \approx 0$.
Furthermore, in this limit, the Jacobian in \eqref{eq:sc_wig} reduces to the stability (or monodromy) matrix for the classical motion, which has unit determinant.
The extension of both these approximations to all values of the final chord $\bm{\xi}'$, over which one integrates in \eqref{eq:sc_wig}, then produces the inverse SFT to \eqref{eq:chord_FT}, that is, the evolved Wigner function is reduced to
\begin{multline}\label{eq:evol_w}
W_{\alpha}(\mathbf{x}',t) \approx \int \frac{\mathrm{d}\xi_{p}\, \mathrm{d}\xi_{q}}{2\pi\hbar} \exp\left[ \frac{i}{\hbar} \bigl( \mathbf{x}(\mathbf{x}',-t) \cdot \mathrm{J}\bm{\xi} \bigr) \right] \chi(\bm{\xi}) \\
= W_{\alpha}\bigl( \mathbf{x}(\mathbf{x}',-t) \bigr) \, ,
\end{multline}
which is just the TWA.

The radical extrapolations that lead to the TWA are not required within the analogous SC approximation for the evolved chord function,
\begin{multline}\label{eq:evol_c}
\chi(\bm{\xi}',t) \approx \int \frac{\mathrm{d}p'\, \mathrm{d}q'}{2\pi\hbar} \left| \det \frac{\mathrm{d}\mathbf{x}}{\mathrm{d}\mathbf{x}'} \right|^{\frac{1}{2}}\\
\times  \exp\left[  -\frac{i}{\hbar} \left( S(\eta_{\pm}',\eta_{\pm}) + \mathbf{x}'\cdot \mathrm{J}\bm{\xi}') \right)\right] W(\mathbf{x}) \, ,
\end{multline}
which is also presented in \cite{ozovz2013}.
One should note that the pair of trajectories, which determine the nonlinear action \eqref{eq:action}, is here defined by a fixed final chord, $\bm{\xi}'$, while we integrate over their final centre, $\mathbf{x}'$; in contrast, it is this centre which is kept fixed in the construction of the appropriate nonlinear action for \eqref{eq:sc_wig}.

The crucial difference between \eqref{eq:sc_wig} and \eqref{eq:evol_c} is that the approximation near the origin supplies a valid local description of the evolved chord function, with no need to incorporate large chords within a rash extrapolation.
In the case of an initial coherent state, $W_{\alpha}(\mathbf{x})$, not only the relevant centres, $\mathbf{x}$, lie within the Gaussian, but the initial chords $\bm{\xi}$ in Fig.~\ref{fig:traj} are also constrained by the Gaussian in \eqref{eq:coh_chord}.
Thus, the choice of small final chords $\bm{\xi}'$ guarantees that the appropriate pair of trajectories will remain near neighbours for all final centers $\mathbf{x}'$ over which one integrates in \eqref{eq:evol_c}.
In this way $S(\eta'_{\pm},\eta_{\pm})\approx 0$ and the Jacobian determinant in the integral can again be ignored.
Indeed, according to \cite{ozovz2013}
\begin{equation}
4 \det \frac{\mathrm{d}\mathbf{x}}{\mathrm{d}\mathbf{x}'} = \det \frac{\mathrm{d}\bm{\xi}}{\mathrm{d}\bm{\xi}'} \xrightarrow{\bm{\xi}'\rightarrow 0 } 1 \, ,
\end{equation}
so that the SC chord function for an evolved coherent state reduces to
\begin{equation}
\chi_{\alpha}(\bm{\xi}',t) \xrightarrow{\bm{\xi}'\rightarrow 0} \int \frac{\mathrm{d}p'\,\mathrm{d}q'}{2\pi\hbar}\,\exp\left[ -\frac{i}{\hbar} \mathbf{x}' \cdot \mathrm{J} \bm{\xi}' \right] \, W_{\alpha}\left( \mathbf{x}(\mathbf{x}',-t)  \right) .
\end{equation}

The truncated chord approximation (TCA) is the SFT of the classical evolution of the Wigner function.
For small chords it coincides with the full SC approximation of the evolved chord function.
It also coincides with the SFT of the TWA, even though the present deduction bypasses the TWA.

The TCA is our main theoretical result.
In the remainder of this paper, the quantum information contained in this small region of the evolved chord function will be investigated numerically.
One should avoid the temptation to put too much substance onto the formal relation
\begin{equation} \label{eq:chord_tca}
    \chi_{\alpha}(\bm{\xi}',t) \xrightarrow{\bm{\xi}'\rightarrow 0} \chi_{\alpha}\bigl( \bm{\xi}(\bm{\xi}',-t) \bigr) \, ,
\end{equation}
in analogy with the TWA \eqref{eq:evol_w}.
It is not clear how to make complete sense of the underlying classical backwards motion $ \bm{\xi}(\bm{\xi}',-t)$, since small chords evolve as tangent vectors for the individual trajectories in the normal phase space.
It is only for large chords that a SC approximation for the chord function determines corresponding centers, $\mathbf{x}'(\bm{\xi}')$, and hence the backwards trajectory, around which the tangent vector evolves linearly \cite{zamo2008}.
Thus, it is more secure to consider the TWA as scaffolding and then take its SFT to obtain the TCA.

If one reverses back to the TWA by taking the full SFT of the TCA, it is clear that it flounders because the evolved chord function is not concentrated on the origin.
The large chords in the SC chord function \cite{zamo2010} account for fine oscillations with wave vector $\bm{\xi}/\hbar$ in the corresponding SC Wigner function \cite{ber1977}, which are killed by the TWA.

In a way, this also explains the failure of the TWA in the deduction from the Gr\"onewold-Moyal formula (see \eg \cite{ozo1998}) for the Weyl-Wigner representation of the commutator of the density operator, $\hat{\rho} = \left| \psi \right>\left< \psi \right|$, with the Hamiltonian:
\begin{multline}
 \bigl[ \,\hat{H},\hat{\rho} \,\bigr](\mathbf{x}) = \frac{i}{\pi\hbar} \sin\left(\frac{\hbar}{2}\frac{\mathrm{d}}{\mathrm{d}\mathbf{x}_{1}} \cdot \mathrm{J}\frac{\mathrm{d}}{\mathrm{d}\mathbf{x}_{2}} \right) \\
 \times H(\mathbf{x}_{1})\big|_{\mathbf{x}_{1}=\mathbf{x}} \, W(\mathbf{x}_{2})\big|_{\mathbf{x}_{2}=\mathbf{x}} \, .
\end{multline}
Apparently, one obtains an attractive decreasing series in powers of $\hbar$, so that the truncation at the first term supplies the TWA.
But this is rendered useless by typically oscillatory waves in a SC Wigner function \cite{ber1977}, with wavelengths of order $\hbar$, since the $n$-th derivative kills off the coefficients $\hbar^{n}$ in the formal series.
This construction of the TWA gives no inkling of the error incurred, even though its failure can again be attributed to the large chords in the classically evolving SC state.
In contrast, the full SC Wigner function is given by \eqref{eq:sc_wig} and the evolving SC chord function is described by \eqref{eq:evol_c}.

\section{Kerr System}\label{sec:IV}

The 4th-order Kerr Hamiltonian was shown in \cite{lanvi2019} to be an excellent test for the SC approximation \eqref{eq:sc_wig} for the evolving Wigner function, far beyond the expected Ehrenfest time. 
It is essentially the square of the Hamiltonian of a simple harmonic oscillator. 
With an appropriate choice of units for position, momentum and energy, the classical Kerr Hamiltonian can be brought into the form
\begin{equation}\label{eq:kerr_ham}
    H(p,q) = \left( p^{2} + q^{2}  \right)^{2} \ .
\end{equation}

\begin{figure}
    \centering
    \includegraphics[width=0.95\linewidth]{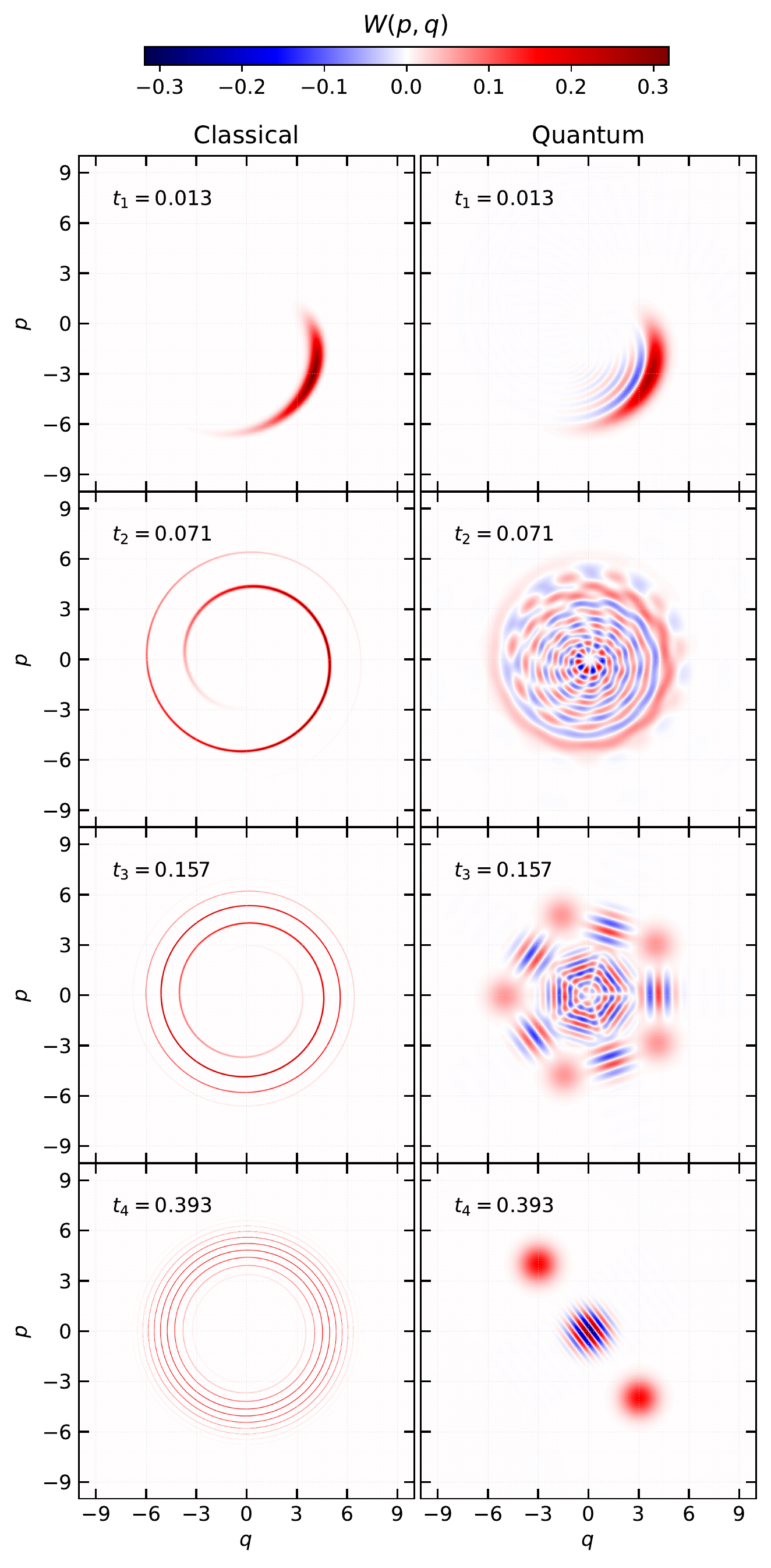}
    \caption{The classical and quantum evolutions of the Wigner function of a coherent state initially centered at the phase-space point $(q,p)=(4,3)$ are displayed for four different times, with the largest one exceeding $6T_{E}$, the Ehrenfest time for this case being given by \eqref{eq:TE} as $T_{E} \approx 0.063$. The times $t_{3} = T_{\mathrm{rev}}/5$ and $t_{4} = T_{\mathrm{rev}}/2$ correspond to fractional revivals, where $T_{\mathrm{rev}}$ is the full revival time. We have set $\hbar=1$.}
    \label{fig:wigner}
\end{figure}

From Hamilton's equations, $H(p,q)$, one finds that
\begin{equation} \label{eq:freq}
    \omega = 4\left( p^{2} + q^{2} \right) \ ,
\end{equation}
is conserved for each orbit, for which it plays the role of an angular frequency in the analytic solution of the equations of motion.
Since orbits with a larger radius have higher angular velocities, the initial classical distribution will both revolve around the origin and stretch into a thin spiraling filament, as can be seen in the evolution of a coherent state displayed in the left column of Fig.~\ref{fig:wigner} for four different times $t$, following \cite{lanvi2019}.

Quantum mechanically, $q$ and $p$ in \eqref{eq:kerr_ham} become operators satisfying the commutation relation $\left[  \hat{q},\hat{p}\right] = i$ (henceforth, we adopt $\hbar =1$).
Introducing the number operator $\hat{n}$ of the harmonic oscillator, we can express the Kerr Hamiltonian as
\begin{equation}\label{eq:QKerr}
    \hat{H} = \left( \hat{p}^2 + \hat{q}^2 \right)^{2} = \left( 2\hat{n} + 1 \right)^{2} \ ,
\end{equation}
with its Fock eigenstates $\left| n \right>$.
The quantum evolution is periodic with the revival time $T_{\text{rev}} = \pi/4$ for the set of units chosen above, before which an initial state undergoes kaleidoscopic partial revivals portrayed in Fig.~\ref{fig:wigner}.
The quantum evolution can be obtained exactly on a grid of fractions of the revival time, which can be chosen to be arbitrarily fine, as reviewed in \cite{lanvi2019}.

The TWA of an initial Gaussian Wigner function for a coherent state driven by the Kerr Hamiltonian is obtained by solving the Liouville equation. 
This was recently picked as an illustration of the non-positivity of the TWA \cite{klisr2020}.
In Fig.~\ref{fig:wigner} we reproduce the classical (left column) and quantum (right column) exact Wigner functions for the evolution of an initial coherent state at four distinct times in \cite{lanvi2019}. 
We see that for $t_{1}$ the classical backbone is clearly visible in the quantum Wigner function, together with the typical interference patterns.
It was shown in \cite{lanvi2019} that the full SC approximation (9) was able to reproduce the exact quantum patterns even for the revivals far beyond the Ehrenfest time
\begin{equation}\label{eq:TE}
    T_{E} = \frac{2\pi}{\omega_{c}} \ ,
\end{equation}
in which the initial distribution's center, moving with angular velocity $\omega_{c}$, has performed a full revolution around the origin \cite{schvt2012}.
For $t_{2}$, for instance, we have already exceeded $T_{E}$ and the multiple interference masks the relationship to the underlying classical spiral structures.
The following panels for $t_{3}$ and $t_{4}$ display fractional revival patterns.
Their time values of $t_3 = T_{\mathrm{rev}}/5$ and $t_4 = T_{\mathrm{rev}}/2$ by far surpass the Ehrenfest time: not only has the classical filament become quite thin, but the gap between different windings of the growing spiral has narrowed.
Obviously it is hopeless to obtain any quantum information for $t_3$ and $t_4$ from the TWA, but its SFT will be shown to reproduce quantum features for the first and the second times in this figure.

\begin{figure*}
    \centering
    \includegraphics[width=0.98\linewidth]{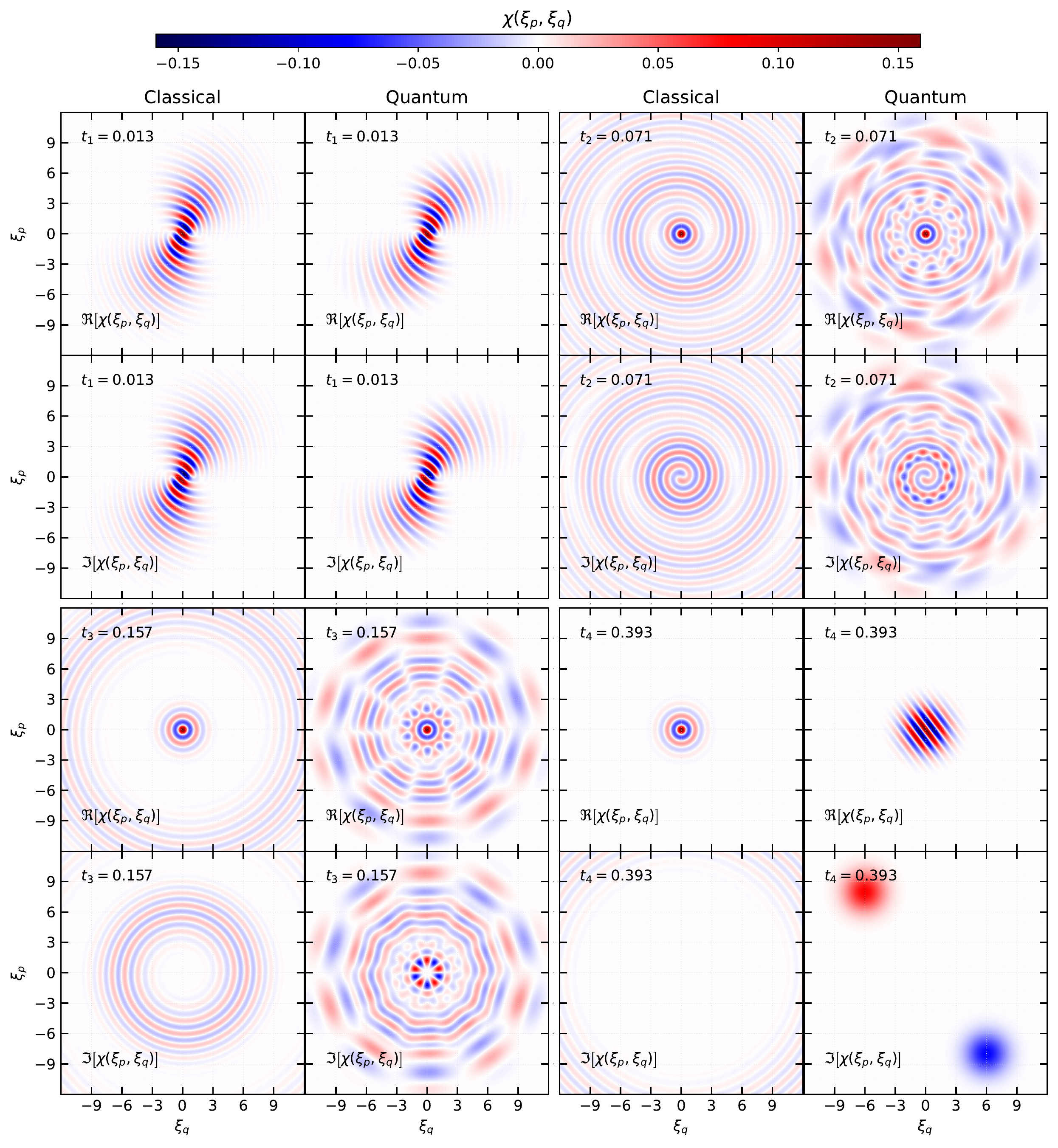}
    \caption{The classical and quantum evolutions of the chord  function of a coherent state initially centered at the phase-space point $(q,p)=(4,3)$ are displayed for the same four different times as in Fig.~\ref{fig:wigner}.}
    \label{fig:chord}
\end{figure*}

\begin{figure}
\centering
\includegraphics[width=0.90\linewidth]{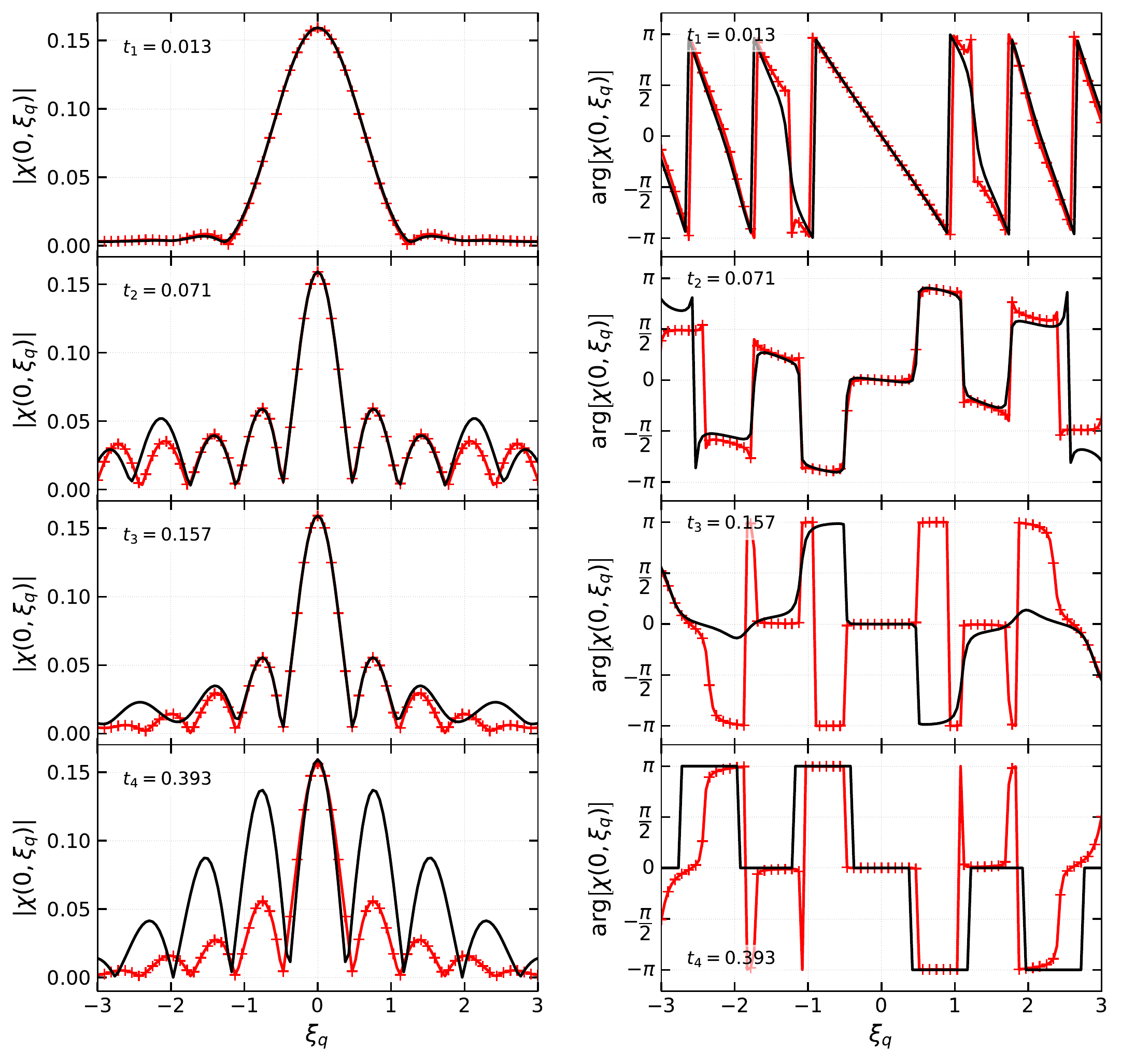}
\caption{The classical and quantum evolution of the chord function of a coherent state initially centered at the phase-space point $(q,p)=(4,3)$ for the same four times as in Fig.~\ref{fig:chord}. Here, we have set $\xi_{p}=0$, and we focus on the region near the origin along $\xi_{q}$. Since the chord function is complex, in the left panel we show the amplitude of the quantum (solid black line) and the classical (crossed red markers) chord functions, while in the right panel the argument of classical and quantum chord functions for the same times are showed. We have set $\hbar=1$.}
    \label{fig:chord_origin}
\end{figure}

This \textit{Wigner picture} of the Kerr evolution is here complemented by its SFT, the \textit{chord picture} presented in Fig.~\ref{fig:chord} for the same evolution times as in Fig.~\ref{fig:wigner}.
The first and third columns show the classical characteristic function split into real and imaginary parts, that is, the SFT of the TWA; whereas the second and fourth columns exhibit the quantum chord function, \ie the chord transform \eqref{eq:chord} of the exact evolution of the same initial coherent state.
For very short times, $t_{1} \approx T_{E}/5$, quantum and classical results for the chord function show similarities in both real and imaginary parts.
However, as time goes on there is no resemblance between classical and quantum results, except in the region near the origin, which is focused in Fig.~\ref{fig:chord_origin}, so as to display the success of the TCA in its domain of applicability.
Since the chord function is, in general, complex, for Fig.~\ref{fig:chord_origin} we have considered it to be written as $\chi(\bm{\xi}) = \left| \chi(\bm{\xi}) \right| \exp\left[ i \arg(\chi(\bm{\xi}))  \right]$.
Thus we compared the amplitude and the phase for the TCA and its quantum counterpart.

Even though the revival as a Schr\"odinger cat occurs at $t_{4}$ in Fig.~\ref{fig:wigner}, a time that lies beyond the applicability of the TCA, this case highlights the complementarity of the Wigner and the chord pictures. 
If this were a symmetric cat state with equal phases for both coherent states, the chord function would be a mere rescaling of the Wigner function \footnote{This holds for all Wigner functions that have a center of symmetry \cite{ozo2009book}.} and, hence, the imaginary part would be uniformly zero. 
In contrast, Fig.~\ref{fig:chord} clearly displays that this is not a symmetric cat state, albeit this fact would only be elicited by a very careful investigation of Fig.~\ref{fig:wigner} of the exact placement of the maxima and minima of the central interference pattern.  

\section{Quantum Blind Spots in the TCA}\label{sec:V}

It is instructive to re-express the chord function \eqref{eq:chord_FT} and \eqref{eq:chord} in the form of an overlap of a pair of states:
\begin{equation}\label{eq:chord_trans}
2\pi\hbar\,\chi(\bm{\xi}) = \left< \psi \right| \hat{T}_{-\bm{\xi}} \left| \psi \right> \, ,
\end{equation}
where
\begin{equation}
\hat{T}_{\bm{\xi}} = \int \mathrm{d}q \, \left| q + \sfrac{\xi_{q}}{2} \right> \left< q - \sfrac{\xi_{q}}{2} \right| \, \ex^{i\xi_{p} q/\hbar}
\end{equation}
is the Weyl operator corresponding to the uniform classical translation by the chord $\bm{\xi}$.
Thus, the correlation of the state portrayed by the chord function with its image translated by $\bm{\xi}$ is 
\begin{equation}
C(\bm{\xi}) = 2\pi\hbar \int \mathrm{d}\mathbf{x}\, W(\mathbf{x}) \, W(\mathbf{x}+\bm{\xi}) = (2\pi\hbar)^{2} \left| \chi(\bm{\xi}) \right|^{2} \, .
\end{equation}

\begin{figure}
    \centering
    \includegraphics[width=0.90\linewidth]{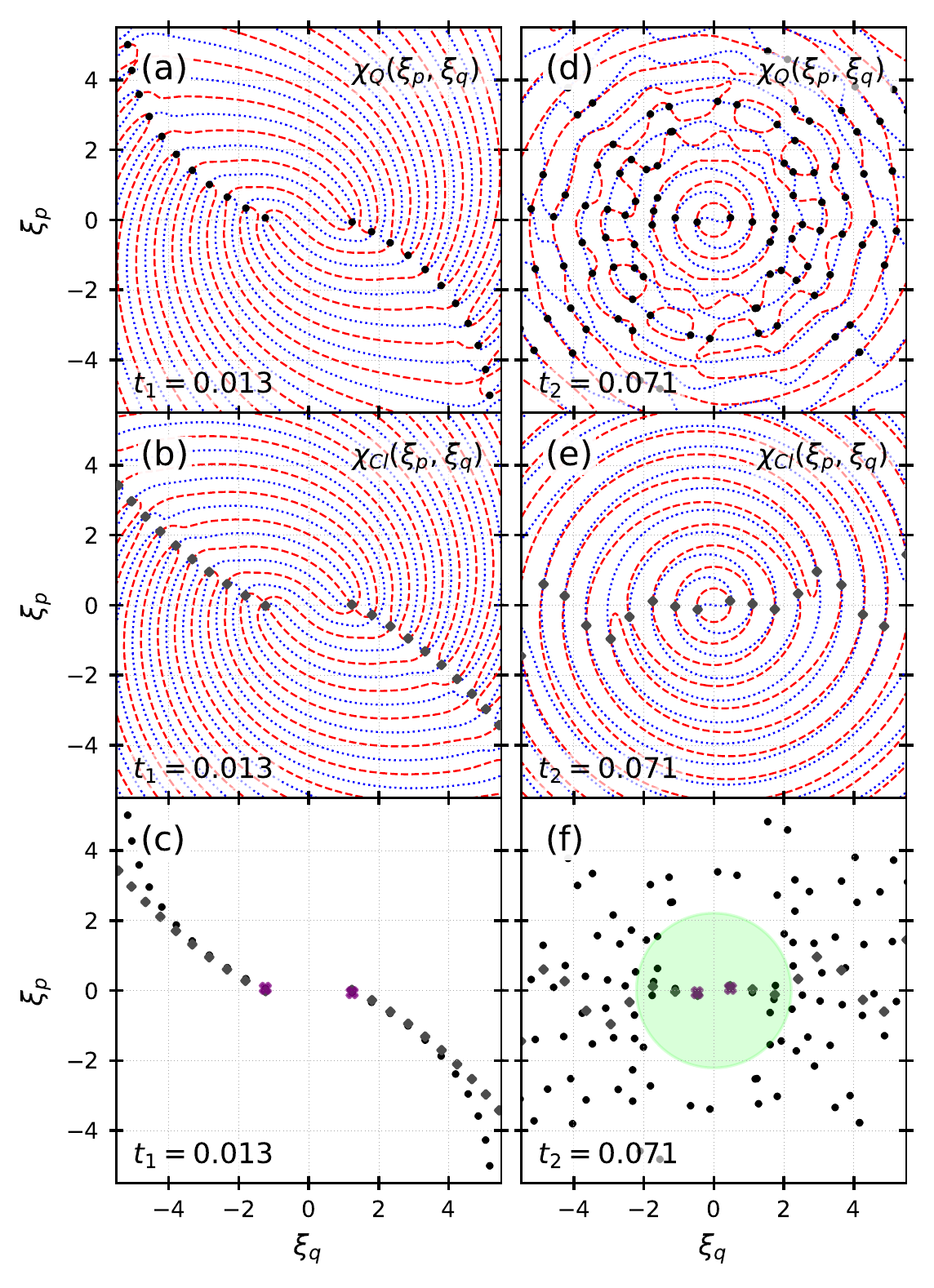}
    \caption{Nodal lines and blind spots for two different times, $t_{1}=0.013$ and $t_{2}=0.071$. Blue dotted lines correspond to nodal lines for the real part of the chord function, while red dashed line indicates the nodals for the imaginary part of the chord function. In (a) and (d), we indicate the blind spots (black dots) for the exact quantum chord function for $t_{1}$ and $t_{2}$, respectively. In (b) and (e), for $t_{1}$ and $t_{2}$ respectively, the blind spots (gray rhombus) for the TCA are shown. Panels (c) and (f) show simultaneously the blind spots obtained for the exact and approximated chord function for $t_{1}$ and $t_{2}$; here the nearest to the origin blind spots are marked with purple crosses. Panel (f) shows a circular area in green determined by the radius $\xi_m$; this is explained and discussed in Appendix \ref{appen}.}
    \label{fig:blind}
\end{figure}

If one considers the Wigner function to be just a probability density (which is always possible for the TWA) the integral above would merely represent the joint probability to find the system in both the distribution $W(\mathbf{x})$ and its translation $W(\mathbf{x}+\bm{\xi})$.
Obviously, for the TWA for the Kerr evolution in Fig.~\ref{fig:wigner}, this will diminish smoothly from $C(0)=1$, becoming negligible only for a translation of the order of the classical spiral, that is $\left| \bm{\xi} \right| > 2\left| \bm{\alpha} \right| $, \ie twice the separation of the original coherent state from the origin.
In contrast, the exact chord function displays \textit{blind spots}, chords at which the chord function cancels \cite{zamo2009}, so that according to \eqref{eq:chord_trans} the translated state $\hat{T}_{-\bm{\xi}} \left| \alpha \right>_{t}$ becomes exactly orthogonal to $\left| \alpha \right>_{t}$.
Those points are determined by the intersection of a nodal line of the real part, $\Re\left[ \chi(\bm{\xi}) \right]=0$, with a nodal line of the imaginary part, $\Im\left[ \chi(\bm{\xi}) \right]=0$.
The blind spots near the origin are marked in Fig.~\ref{fig:blind} and perhaps surprisingly they also show up in the corresponding TCA which thus affords an approximate placement of these purely quantum features.

The special ingredient of the present investigation of blind spots is that they come in a continuous family, parameterized by time of the evolution. In terms of the initial state they concern the zeroes of
\begin{equation}
L(\bm{\xi},t) = \left<\alpha \right|  \hat{U}_{-t}\, \hat{U}'_{t}(\bm{\xi}) \left| \alpha \right> \, ,
\end{equation}
where $\hat{U}_{t}$ is the evolution operator in \eqref{eq:evol} and its slight variation is $\hat{U}'_{t}(\bm{\xi}) = \hat{T}_{\bm{\xi}}\, \hat{U}_{t}$.
For each value of the two dimensional continuous parameter $\bm{\xi}$, this is just \textit{the quantum Loschmidt echo}, \ie the evolving \textit{fidelity} for a pair of evolutions.
It was proposed by Peres \cite{per1984} as a measure of quantum chaos and its relation to the classical Lyapunov exponent for chaotic systems was first verified in \cite{jalp2001} (reviewed in \cite{goujp2012}).

Even though our computations do not deal with a chaotic state, it is shown in \cite{zamo2010} that blind spots are a generic feature of pure states.
Hence, they are present in states resulting from general evolutions.
It follows that, at least for variations obtained from translations, the choice of the parameter $\bm{\xi}$ is of crucial importance, since one obtains an exact zero of the fidelity, instead of a smooth decay.
The latter emerges only through some kind of averaging of the displacement of the initial state.

\section{Expectations of Observables}\label{sec:VI}

The Weyl representation for a common mechanical observable $\hat{O}$, \ie
\begin{equation}
O(\mathbf{x}) = \int \mathrm{d}\tilde{q}\, \bigl< q+\tilde{q} \bigr| \, \hat{O} \, \bigl| q - \tilde{q} \bigr>\, \ex^{-2ip \tilde{q}/\hbar} \, ,
\end{equation}
is just the corresponding classical phase space function, or very close to it.
In contrast, the division by $\hbar$ in the exponent for its SFT, just as in \eqref{eq:chord_FT}, pushes its chord representation
\begin{equation}
\tilde{O}(\bm{\xi}) = \int \frac{\mathrm{d}\mathbf{x}}{2\pi\hbar}\, \exp\left[ -\frac{i}{\hbar} \mathbf{x}\cdot\mathrm{J} \bm{\xi} \right] \, O(\mathbf{x}) \, ,
\end{equation}
within a classically small neighborhood of the chord origin.
Thus, in spite of the similar integrals for the expectation of an observable in both these representations,
\begin{equation} \label{eq:expec_t}
\bigl< \hat{O} \bigr> = \int\mathrm{d}\mathbf{x}\, O(\mathbf{x})\,W(\mathbf{x},t) = \int \mathrm{d}\bm{\xi}\, \tilde{O}(\bm{\xi})\, \chi(-\bm{\xi},t) \, ,
\end{equation}
they provide complementary pictures: while the Wigner function evaluates the expectations as if it were a classical average (Wigner's original motivation), it is only the central part of the chord function that matters in the alternative calculation.
Extreme examples are provided by the exact expressions for moments:
\begin{align}
\left< \hat{q}^{n} \right> &= 2\pi \hbar\, (i\hbar)^{n} \, \frac{\mathrm{d}^{n} \chi}{\mathrm{d}\xi_{p}^{n}} \Bigg|_{\bm{\xi}=0} \, , \\
\left< \hat{p}^{n} \right> &= 2\pi\hbar \,  (-i\hbar)^{n} \, \frac{\mathrm{d}^{n} \chi}{\mathrm{d}\xi_{q}^{n}} \Bigg|_{\bm{\xi}=0} \, .
\end{align}

Thus, the outlying regions of the chord function play no role in these expected values, so that the time evolution of expectations such as $\left< \hat{p}^{n} \right>_{t}$ calculated from \eqref{eq:expec_t} with the approximated chord function $\chi(\bm{\xi}',t)$ given by the TCA \eqref{eq:chord_tca}, should give reasonable estimates.
But if the second integral in \eqref{eq:expec_t} is reliable, then the exact SFT of both factors within this integral, inserted into the first integral, will give the same result.
Therefore, the TWA $W(\mathbf{x}(\mathbf{x}',-t))$ also provides a good estimate of the expectation, which accounts somewhat for its paradoxical popularity.

\begin{figure}
    \centering
    \includegraphics[width=0.99\linewidth]{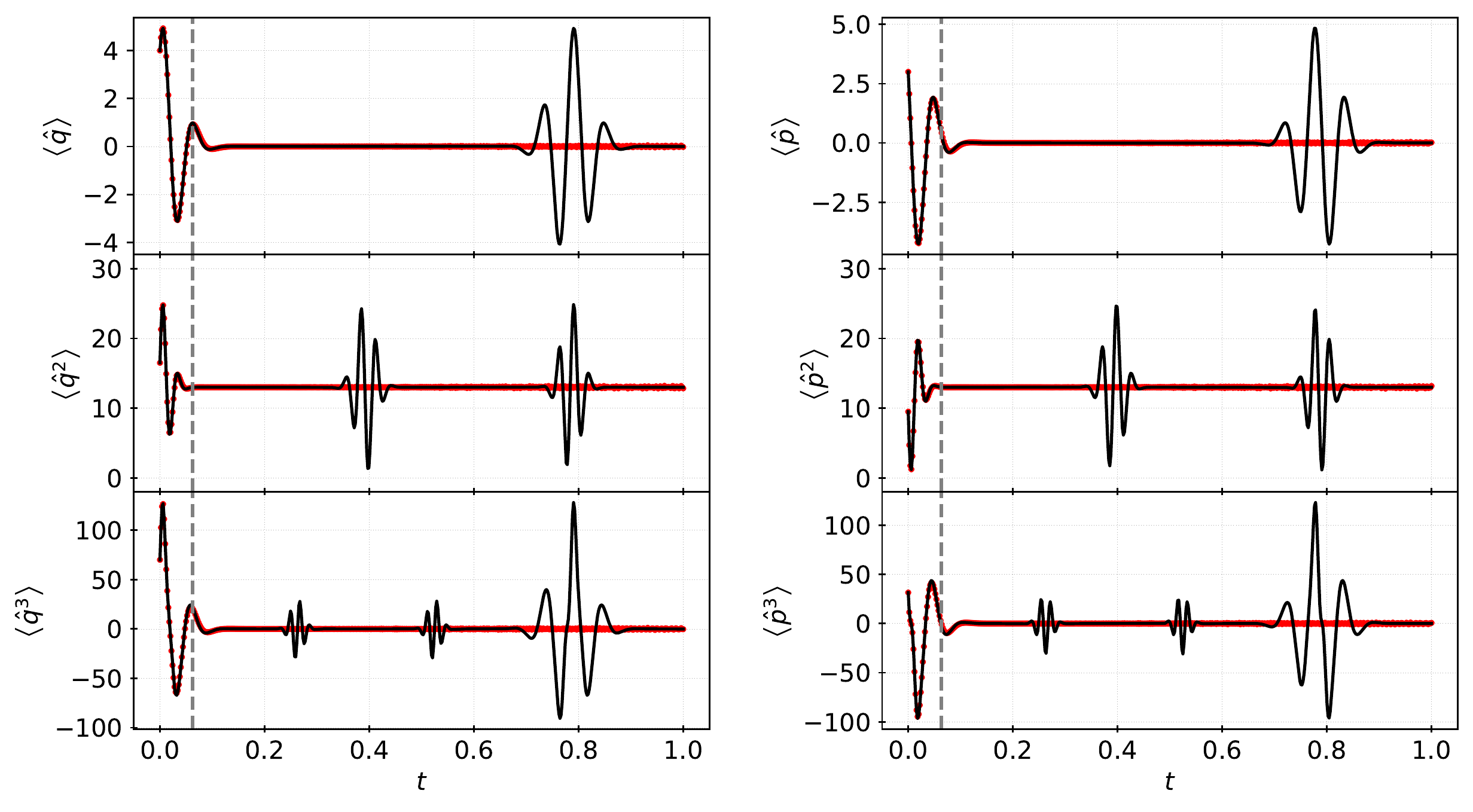}
    \caption{Comparison of the moments for $\hat{p}$ and $\hat{q}$ for the initial coherent state $\left| \alpha \right>$, with $\alpha = \left(4+3i \right)/\sqrt{2}$, under the dynamics ruled by the Kerr Hamiltonian. Solid black lines corresponds to exact quantum calculation, while red points are those calculated via the TWA. Vertical gray dashed lines indicate the Ehrenfest time $T_{E}$ for the system.}
    \label{fig:expect}
\end{figure}

Fig.~\ref{fig:expect} compares the moments calculated through the TWA in \eqref{eq:expec_t} with the exact quantum moments up to 3rd-order in $\hat{p}$ and $\hat{q}$, there we observe the agreement between the two calculations up to the Ehrenfest time $T_{E}$.
For longer times the classical approximation does not recover the quantum behavior.

\section{Local Correlations}\label{sec:VII}

In the context of integrable and ergodic wave functions \cite{ber1977a}, local correlations of the wave function were shown by Berry to exhibit quantum features even in an apparently pure classical approximation.
Originally, a sharp window of width $\Delta$ limited the \textit{classically small region} over which the correlation is evaluated, but a Gaussian window will also do as in \cite{zamzo2015}, so that the local correlation is
\begin{multline}\label{eq:bwflc}
\mathbf{C}_{\Delta}(\xi_{q},Q) \equiv \frac{1}{\nu_{\Delta}(Q)} \int \frac{\mathrm{d}q}{\sqrt{2\pi}\Delta} \, \ex^{-\sfrac{(q-Q)^{2}}{2\Delta^{2}}}\\
\left< q + \sfrac{\xi_{q}}{2} \right| \left. \psi \right> \left< \psi \right|\left. q - \sfrac{\xi_{q}}{2} \right> \, ,
\end{multline}
where $\nu_{\Delta}(Q)$ is the normalization factor such that $\mathbf{C}_{\Delta}(0,Q) =1$.

Still following \cite{zamzo2015}, the local wave function correlation can be re-expressed as an integral over the chord function,
\begin{equation}\label{eq:bwflc_2}
\mathbf{C}_{\Delta}(\xi_{q},Q) = \frac{1}{\nu_{\Delta}(Q)} \int \mathrm{d}\xi_{p} \: \chi(\bm{\xi}) \: \ex^{\frac{i}{\hbar}\xi_{p}\, Q} \: \ex^{-\frac{\Delta^{2}}{2\hbar^{2}}\xi_{p}^{2}} \, ,
\end{equation}
with the normalization evaluated as
\begin{equation}
\nu_{\Delta}(Q) = \int \mathrm{d}\xi_{p} \: \chi(\xi_{p}, 0) \: \ex^{\frac{i}{\hbar}\xi_{p}\, Q} \: \ex^{-\frac{\Delta^{2}}{2\hbar^{2}}\xi_{p}^{2}} \, .
\end{equation}

\begin{figure}
    \centering
    \includegraphics[width=0.98\linewidth]{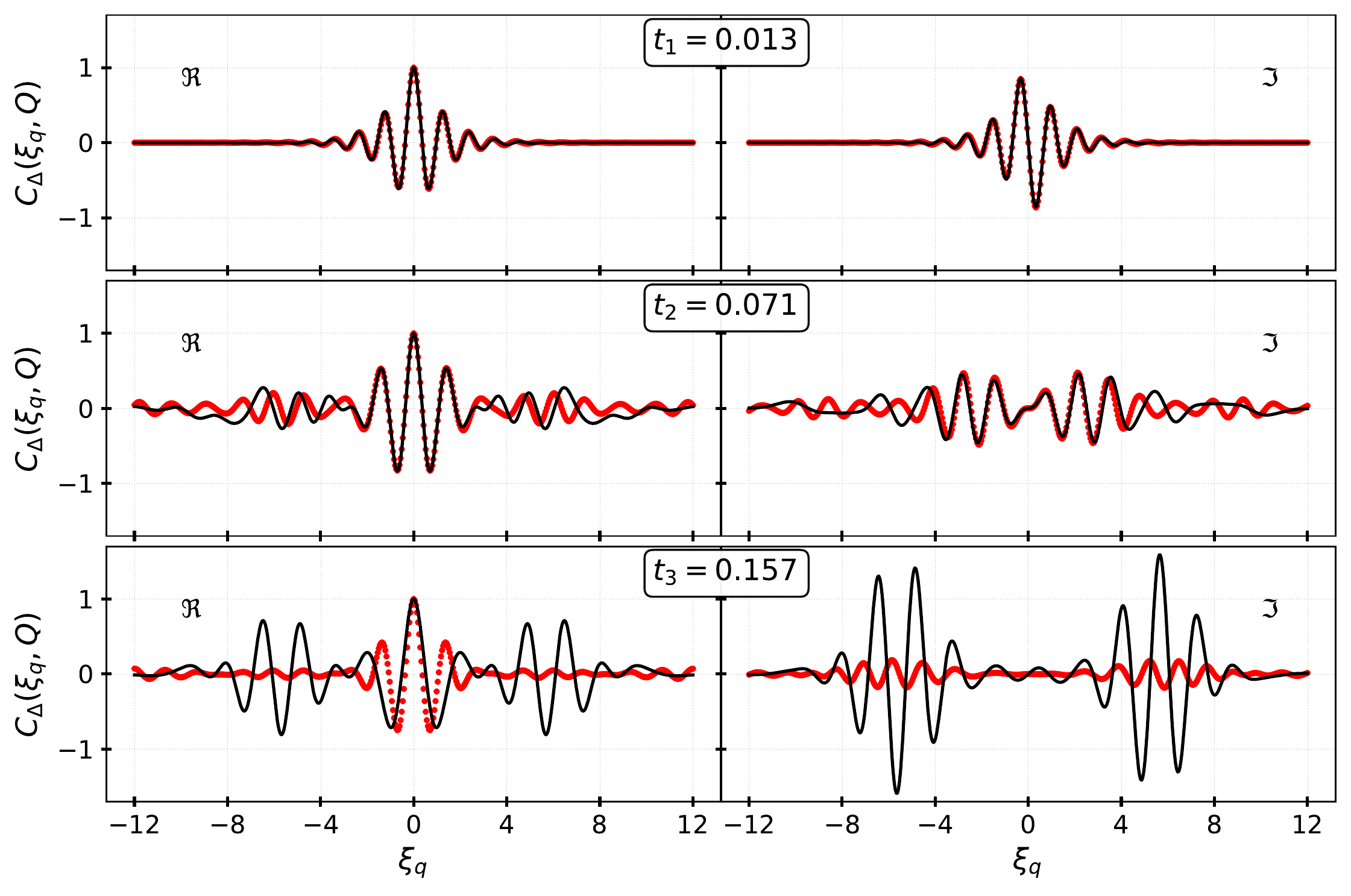}
    \caption{Local correlations of the wave function computed using the TCA (red dots) and the exact quantum chord function (solid black line) of a coherent state, initially centered at the phase space point $(q,p)=(4,3)$, evolving under the action of the Kerr Hamiltonian. The left panel shows the real part of the correlation, while the right panel shows the imaginary part. The comparison is made for the first three times in Fig.~\ref{fig:wigner}. We have set $\Delta=1$, $Q=2$ and $\hbar=1$.}
    \label{fig:corr}
\end{figure}

But if we choose $\Delta \approx \hbar^{1/2}$, then $\xi_{q}$ is small, because of the Gaussian in \eqref{eq:bwflc}, while $\xi_{p}$ is likewise limited by the Gaussian in \eqref{eq:bwflc_2}.
Therefore, the local wave function correlations are basically determined by the chords within a circle of area $\hbar$.
Hence, it is legitimate to approximate the time evolution of these correlations with the TCA in \eqref{eq:bwflc_2}.
The comparison of this approximation with the exact correlation for the Kerr evolution of a coherent state is made in Fig.~\ref{fig:corr}.

\section{Conclusions}\label{sec:conclu}

The general principle by which the central region of the chord function encodes all the information about the classical Liouville distribution corresponding to a quantum state was previously supported by computations on the eigenstates of a system for a chaotic Hamiltonian \cite{zamzo2015}.
Here we go a step further by supplying a direct deduction of the validity of the classical truncated chord approximation (TCA) near the origin from semiclassical formulae \cite{ozovz2013}. 
Indeed, the latter has been shown to reproduce detailed interference patterns in the exact Wigner function \cite{lanvi2019}.
The TCA is here computationally verified to reproduce distinctive quantum features of a coherent state evolved by the quartic Kerr Hamiltonian.

One may wonder about the generality of conclusions based on as simple and special a system as the Kerr. 
Actually, the scenario for classical evolution of a compact region of phase space was classified by Berry et al. \cite{berbt1979} into {\it whorls} and {\it tendrils}, being that the evolved spiral in the present case is merely a pretty example of a well-behaved whorl. 
This the norm for integrable systems and hence any neighbourhood of a stable equilibrium for a single degree of freedom. 
In the \ref{appen} we have explained the separation of the blind spots into an inner {\it classical region} and a seemingly random outer region beyond the reach of the TCA, on the basis of the decreasing separation of the spiral windings. 
The existence of these distinct zones can probably be extended to general whorls.

On the other hand, tendrils are developed by continuous Hamiltonian systems that may be more or less chaotic for two or more degrees of freedom, as well as by the maps studied in \cite{berbt1979}. 
The classical nature of the blind spots for bound states of a chaotic Hamiltonian close to the chord origin was established in \cite{zamzo2015}. 
It is stimulating that a recent paper by Mittal et al. \cite{mitkgu2020} deals with corrections beyond TWA to expectation values of chaotic systems. 
These corrections are attributed to the close return of invariant manifolds (returning tendrils) and hence to small chords just as here, though no mention is made of the chord function.

Notwithstanding its crucial role in the clarification of the delicate validity of fully classical methods for approximating strictly quantum evolution, we in no way advocate the substitution of the classically evolved Wigner function by direct use of the TCA.
This representation is not as intuitive as the TWA, but it is a sure guide to the validity of the approximate expectation for any operator.
If the function that represents this operator in the chord representation is not concentrated near the origin, the result incurs the risk of being completely fictitious.

\begin{acknowledgments}
KT is thankful to the Centro Brasileiro de Pesquisas Fisicas for the warm hospitality and acknowledges the financial support from TWAS-CNPq, together with PIFI-CAS grant No.~2019PM0109.
AMOA gratefully acknowledges partial financial support from CNPq and the National Institute for Science and Technology: Quantum Information.
\end{acknowledgments}

\appendix
\section{Semiclassical interpretation for the layout of blind spots}\label{appen}

In order to understand the general layout of the nearest blind spots to the origin for the evolution of a coherent state by the Kerr system, one should recall two basic SC results. 
The first is the verification in \cite{mairnv2008} that the localized Gaussian distribution in phase space, which evolves from an initial coherent state, condenses onto a thin filament that can be treated as a smooth curve. 
This is the backbone for the various extensions of WKB approximations for static or dynamical states. 
Even though the authors directed their attention to chaotic evolution, it is evident that the classical spiral resulting from the Kerr evolution is another example of this phenomenon.

The other result is that the SC chord function $\chi(\bm{\xi})$, corresponding to such a curve, is a superposition of all the realizations of the given argument $\bm{\xi}$ as geometrical chords of this classical curve \cite{zamo2008}. 
The phase for each realization depends on a classical action which includes the area between the curve and the chord, as in the well known construction of the SC Wigner function \cite{ber1977}. 
A centro-symmetric curve is anomalous, since it corresponds to a real chord function, so that the zeroes of the chord function are not isolated. 
However, in \cite{zamo2010} it was shown that a small deformation of the circle, corresponding to the eigenstate of the harmonic oscillator, supports a chord function with blind spots along straight lines, which radiate from the origin. 

\begin{figure}
    \centering
    \includegraphics[width=0.90\linewidth]{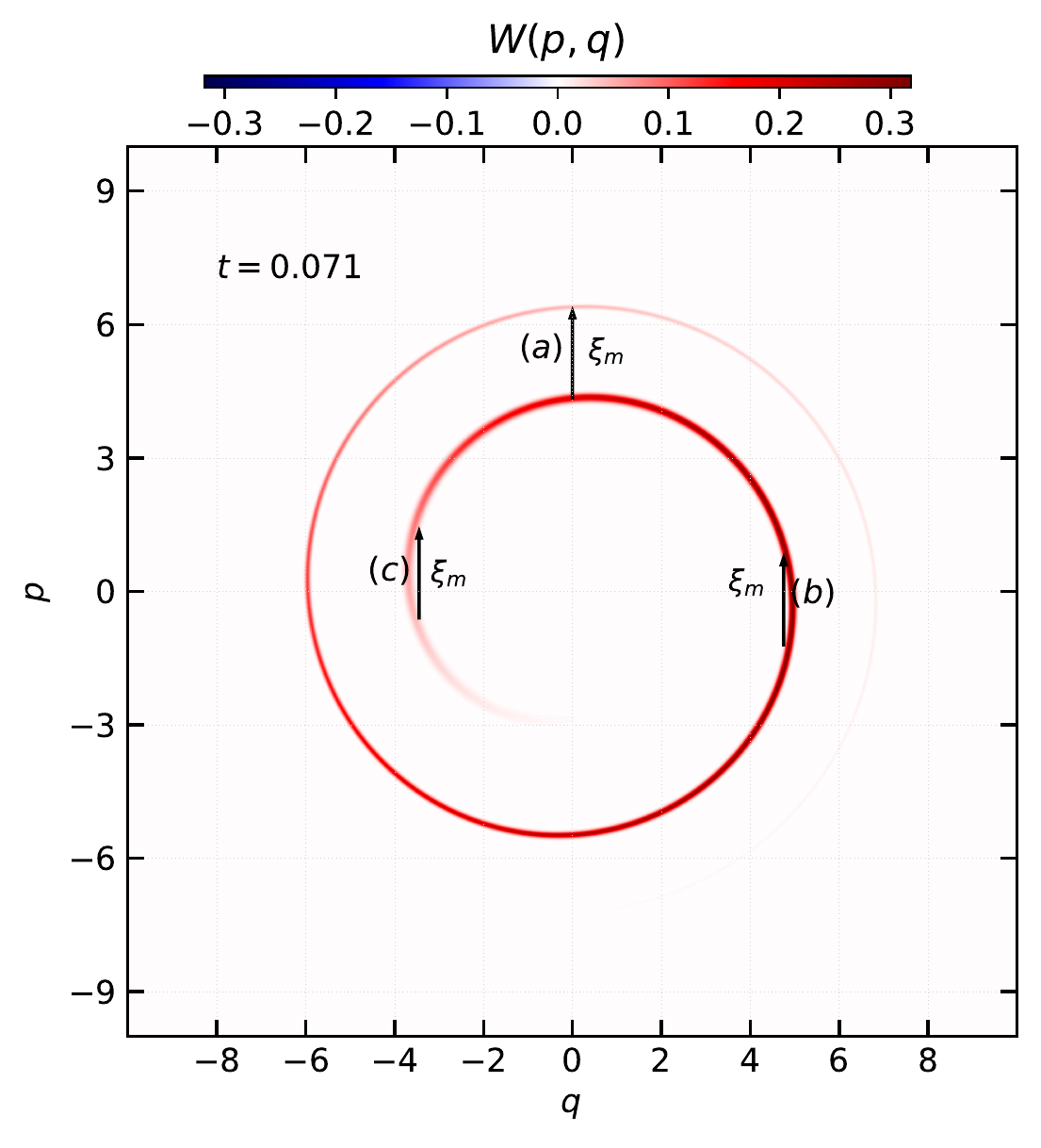}
    \caption{Classical Wigner function at time $t=0.071$ of a coherent state initially centered at the phase-space point $(q,p) = (4,3)$ evolving under the Kerr Hamiltonian. The minimum chord between nearest chords $\xi_{m}(t)$ is shown.}
    \label{fig:B1}
\end{figure}

\begin{figure}
    \centering
    \includegraphics[width=0.95\linewidth]{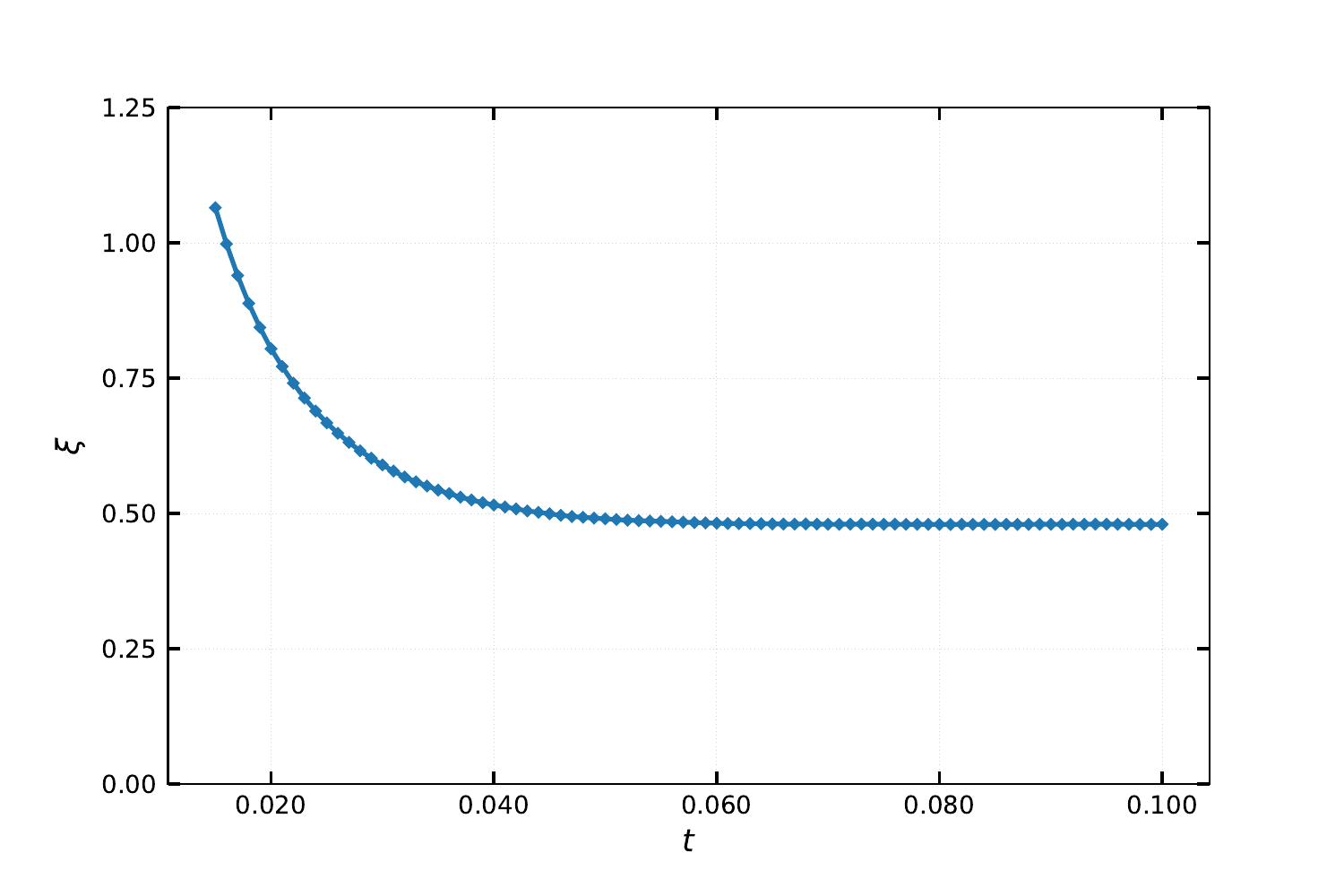}
    \caption{Distance of the nearest blind spot to the origin in the dual phase space as a function of time for an initial coherent state centred at $(q,p)=(4,3)$ evolving under the action of the Kerr Hamiltonian.}
    \label{fig:B2}
\end{figure}

In the present case, instead of closing, the curve describes a spiral that tightens in time. 
Ehrenfest time brings in a new winding, though this is loosely defined by the arbitrary boundary established for the Gaussian representation of the initial coherent state.
Beyond the Ehrenfest time, new chords arise, which join the different windings. This is shown in Fig.~\ref{fig:B1}; defining $\xi_{m}(t)$ as the modulus of the minimum chord between nearest windings at the time $t$, then the SC chord function within the circle of radius $\xi_m$ is only affected by the local branch of the spiral. 
The only change in this region near Ehrenfest time is that a small open curve has only a single realization for each chord, whereas a second realization (as in a closed curve) arises beyond this threshold. 
This is the case of vertical chords smaller than $\xi_m$, near its realizations (b) and (c) in Fig.~\ref{fig:B1}. 
In any case, the phase for each of these realizations depends on the small area between these geometrical chords and the spiral. 
The contribution of each of these local chords is not affected by the passage of Ehrenfest time and one may attribute to them the line of blind spots with a radial tangent, akin to one of the lines for the distorted closed curve in \cite{zamo2008}. 
Indeed, it is shown in Fig.~\ref{fig:B2} that the distance of the nearest blind spot to the origin stabilizes in time as the thickness of the spiral strip can truly be neglected.

\begin{figure}
    \centering
    \includegraphics[width=0.95\linewidth]{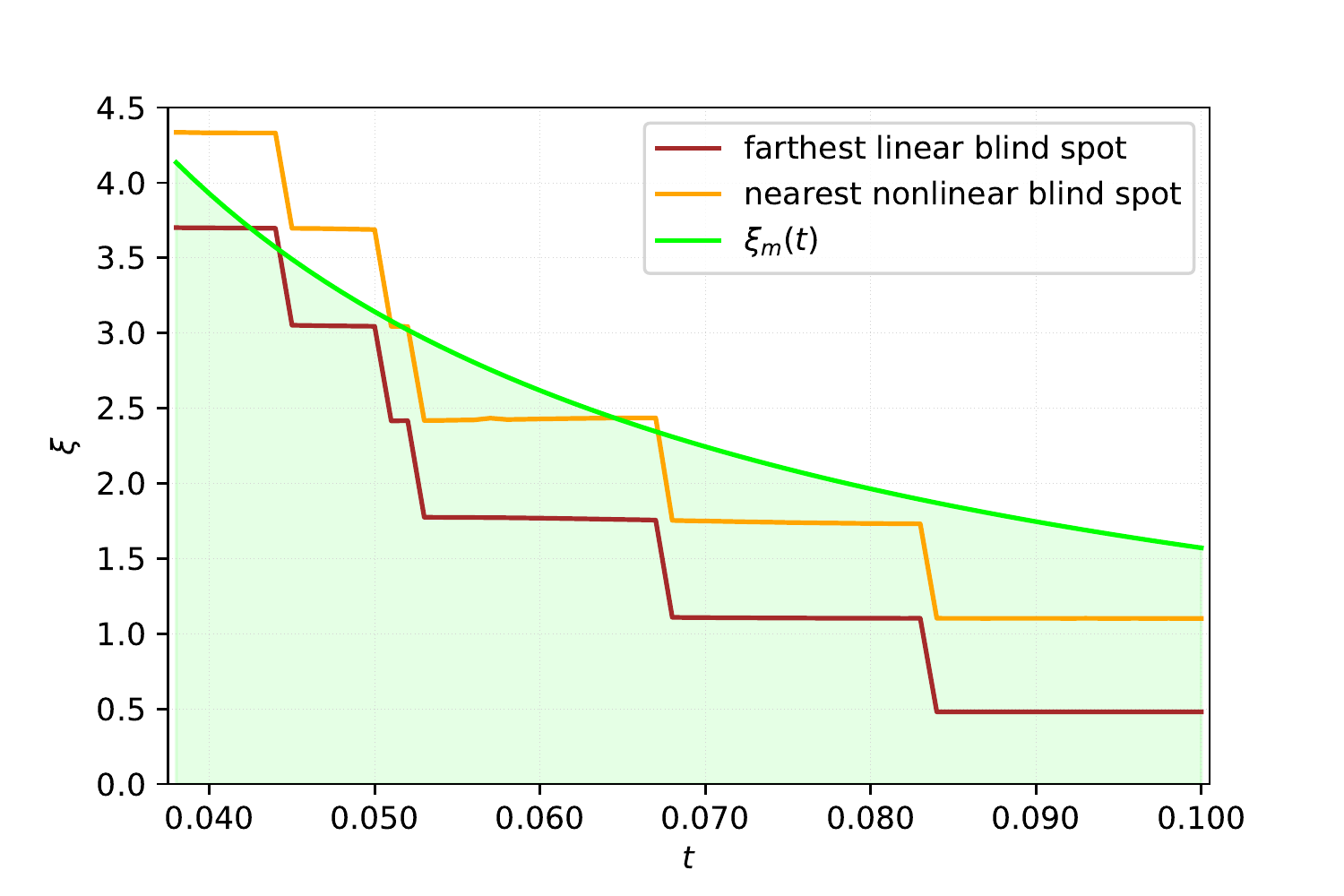}
    \caption{Distance from the origin of the dual chord  phase space of the blind spots near the threshold circle of radius $\xi_{m}(t)$ as a function of time for an initial coherent state centred at $(q,p)=(4,3)$ evolving under the Kerr Hamiltonian. The green line depicts the modulus of the minimum chord between nearest windings $\xi_{m}(t) = \pi/(4tX)$, which corresponds to the estimated radius of the threshold circle. The orange line shows the distance from the origin to the nearest nonlinear blind spot. The brown line indicates the distance from the origin to the farthest linear blind spot.}
    \label{fig:B3}
\end{figure}

The qualitative effect of the new windings beyond the Ehrenfest time appears in the chord function outside the circle of radius $\xi_m$, which is determined by the realization (a) in Fig.~\ref{fig:B1}, \ie the smallest of a new family of vertical chords between both windings. 
These realizations define large areas (and hence, SC actions) with respect to the spiral, adding up to a chord function that has no relation to the Fourier transform of the classical spiral. 
The result is an apparently random scattering of blind spots beyond a time dependent threshold, see Fig.~\ref{fig:blind}: (f). 
This can be estimated for the Kerr system by recalling that the angular velocity along the circular trajectory of an initial point with square radius $x^2 = (p^2 +q^2)$ is $\omega(x) = 4 x^2$. 
Then $\mathrm{d}\omega/\mathrm{d}x = 8 x$ and, for a pair of initial points separated by a radial chord $\xi$, the approximate difference in angle traversed is $8 t x \xi$. 
For the minimum chord of a different winding, this angle is $2\pi$, so that approximately $\xi_m(t) = \pi / (4tX)$, where X is the radius of the centre of the initial coherent state. 
Figure \ref{fig:B3} exhibits the disposition of the farthest linear blind spot, the nearest nonlinear blind spots to the chord origin, and the threshold circle of radius $\xi_m(t)$. 
This confirms that the closest linear blind spots have a classical origin and their number only decreases because the distance between the spiral windings shrinks with time.

%%%%%% BIBLIOGRAPHY %%%%%
%\bibliography{references}{}
%\bibliographystyle{apsrev4-2}
%
\end{document}